\documentclass[a4paper, 12pt]{article}
\usepackage[english]{babel}
\usepackage{sidecap}
\usepackage{textcomp}
\usepackage{graphicx}
\usepackage[colorinlistoftodos]{todonotes}
\usepackage[utf8x]{inputenc}
\usepackage[T1]{fontenc}
\usepackage[round]{natbib}
\usepackage{authblk}
\usepackage{siunitx}
\usepackage{amsmath,amssymb}
\usepackage{mathptmx}
\usepackage{cancel}
\usepackage[onehalfspacing]{setspace}

\title{Active subthreshold dendritic conductances shape the local field potential}
\author[1]{Torbj\o rn V Ness}
\author[2]{Michiel W H Remme}
\author[1, 3\footnote{Corresponding author: gaute.einevoll@nmbu.no}]{Gaute T Einevoll} 
\affil[1]{Department of Mathematical Sciences, Norwegian University of Life Sciences\\ \AA s, Norway}
\affil[2]{Institute for Theoretical Biology, Humboldt University Berlin\\Berlin, Germany}
\affil[3]{Department of Physics, University of Oslo\\Oslo, Norway}
\begin{document}
\mbox{}\\
\textbf{Title:} 

Active subthreshold dendritic conductances shape the local field potential
\\\textbf{Abbreviated title:} 

Active dendritic conductances shape the LFP
\\\textbf{Author names and affiliation:}

Torbj\o rn V Ness,
Department of Mathematical Sciences and Technology, Norwegian University of Life Sciences, 1432 \AA s, Norway.

Michiel W H Remme, Institute for Theoretical Biology, Humboldt University Berlin, 10115 Berlin, Germany.

Gaute T Einevoll, Department of Mathematical Sciences and Technology, Norwegian University of Life Sciences, 1432 \AA s, Norway;
Department of Physics, University of Oslo, 0316 Oslo, Norway.\\
\\\textbf{Corresponding author:}
\\Gaute T Einevoll, Department of Mathematical Sciences and Technology, Norwegian University of Life Sciences, 1432 \AA s, Norway; gaute.einevoll@nmbu.no\\
\\


\pagebreak

\date{}
\maketitle

\onehalfspacing

\subsection*{Key points summary}
\begin{itemize}
\item The local field potential (LFP), the low frequency part of extracellular potentials recorded in neural tissue, is often used for probing neural circuit activity. Interpreting the LFP signal is difficult, however. 
   
\item While the cortical LFP is thought to mainly reflect synaptic inputs onto pyramidal neurons, little is known about the role of the various subthreshold active conductances in shaping the LFP.

\item By means of biophysical modeling we obtain a comprehensive qualitative understanding of how the LFP generated by a single pyramidal neuron depends on type and spatial distribution of active subthreshold currents.

\item For pyramidal neurons, the h-type channels likely play a key role and can cause a distinct resonance in the LFP power spectrum.

\item Our results show that the LFP signal can give information about the active properties of neurons and imply that preferred frequencies in the LFP can result from those cellular properties instead of, e.g., network dynamics.
\end{itemize}

\pagebreak
\begin{abstract}	
The main contribution to the local field potential (LFP) is thought to stem from synaptic input to neurons and the ensuing subthreshold dendritic processing. The role of active dendritic conductances in shaping the LFP has received little attention, even though such ion channels are known to affect the subthreshold neuron dynamics. Here we used a modeling approach to investigate the effects of subthreshold dendritic conductances on the LFP. Using a biophysically detailed, experimentally constrained model of a cortical pyramidal neuron, we identified conditions under which subthreshold active conductances are a major factor in shaping the LFP. We found that particularly the hyperpolarization-activated inward current, $I_{\rm h}$, can have a sizable effect and cause a resonance in the LFP power spectral density. 
To get a general, qualitative understanding of how any subthreshold active dendritic conductance and its cellular distribution can affect the LFP, we next performed a systematic study with a simplified model.
We found that the effect on the LFP is most pronounced when (1) the synaptic drive to the cell is asymmetrically distributed (i.e., either basal or apical), (2) the active conductances are distributed non-uniformly with the highest channel densities near the synaptic input, and (3) when the LFP is measured at the opposite pole of the cell relative to the synaptic input. In summary, we show that subthreshold active conductances can be strongly reflected in LFP signals, opening up the possibility that the LFP can be used to characterize the properties and cellular distributions of active conductances.
\end{abstract}


\subsection*{Abbreviations list}
LFP, Local Field Potential; PSD, Power Spectral Density

\section*{Introduction}

The local field potential (LFP), the low-frequency part ($\lesssim$ 500~Hz) of the extracellular potential recorded in the brain, has experienced a rejuvenated interest in the last decade. This is partly due to the development of novel silicon-based microelectrodes with tens or hundreds of electrode contacts covering large volumes of brain tissue, the realization that the LFP offers a unique window into neural population activity, and the possibilities offered by the signal in steering neuroprosthetic devices~\citep{Buzsaki2012, Einevoll2013}. The interpretation of the LFP signal in terms of neural activity is challenging, however, as in general thousands of neurons contribute to the recorded signal~\citep{Linden2011}. Careful mathematical modeling is thus needed to take full advantage of the opportunities that the LFP signal offers~\citep{Einevoll2013}. 

Fortunately, the biophysical origin of the LFP signals seems well understood in the context of volume conduction theory~\citep{Nunez2006}, and a forward-modeling scheme linking transmembrane neural currents to recorded electrical potentials, including LFPs, has been established~\citep{Rall1968a,Holt1999}. The forward-modeling scheme has been successfully used to study the cortical LFP signal generated by single neurons \citep{Linden2010} as well as cell populations~\citep{Pettersen2008a,Linden2011,Leski2013,Tomsett2015}. These studies showed, for example, that the lower frequencies of the synaptic input current give the largest contributions to the LFP for two reasons. First, the lowest frequencies give the largest transmembrane current dipole moments due to the filtering properties of dendrites~\citep{Linden2010}. Second, increased correlations in synaptic input to a population of neurons can strongly amplify the LFP signal~\citep{Linden2011,Einevoll2013,Einevoll2013a}, and this amplification is strongest for the lowest input frequencies~\citep{Leski2013}. Further, the amplification of the LFP due to synaptic input correlations requires the input to be asymmetrically distributed over the dendrites~\citep{Linden2011,Leski2013}. 

However, these studies were typically carried out using neuron models with passive dendrites. Neurons express a variety of active conductances that are distributed in specific ways across both soma and dendrites~\citep{Stuart2007}. For example, the h-type current is most densely concentrated in the distal apical dendrites \citep{Williams_2000,Magee1998}, whereas the M-type slow potassium current is mostly localized close to the soma \citep{Hu_2007}. It is well established that such active conductances affect the subthreshold processing of synaptic input (see, e.g., \citealt{Remme2011, Major2013}). This is highly relevant for understanding the LFP signal, since, in cortex, the LFP is thought to mainly reflect synaptic currents and their associated return currents~\citep{Mitzdorf1985,Einevoll2007,Einevoll2013}.

Here we use a modeling approach to investigate specifically the role of voltage-dependent conductances in shaping the LFP in the subthreshold regime. We focused on pyramidal cells since they are thought to be the main contributor to the LFP in cortical recordings~\citep{Einevoll2013} and used the state-of-the-art cortical layer 5 pyramidal neuron model by~\citet{Hay2011} as a starting point for our investigation. With this highly detailed model we established that subthreshold active conductances can indeed have a sizable impact on the LFP signal. Using a simplified neuron model that used generalized descriptions of active conductances, so-called quasi-active currents \citep{Koch1984,Remme2014}, we were able to obtain an understanding of how any active conductance can affect the LFP, depending on the characteristics of the current and its distribution across the neuron.

\section*{Methods}

\subsection*{Pyramidal neuron model}
Numerical simulations for Figures 1 and 2 were carried out using a model of a cortical layer 5 pyramidal cell that was published by \cite{Hay2011}. This model has a detailed morphology and includes ten active ionic conductances fitted to experimental data by multi-objective optimization with an evolutionary algorithm. For simulations of a passive model, the active conductances were removed from the model. Simulations with a so-called "frozen" h-type conductance were performed by keeping the gating variable of the h-current constant, yielding an additional passive conductance. 
The original model showed a voltage gradient from soma to distal apical dendrites. In order to simplify the interpretation of the simulation results, we adjusted the leak reversal potential of each compartment such that we could set the resting potential uniformly to a specified chosen potential \citep{Carnevale2006}.

The neuron model received either excitatory synaptic input or white-noise current input. Synaptic inputs were modeled as steps in the synaptic conductance followed by an exponential decay with a time constant of 2~ms and used a reversal potential of 0~mV. The white-noise current input consisted of a sum of sinusoids with identical amplitudes but random phases for each integer frequency from 1~Hz to 500~Hz (see \citealt{Linden2010}). The resulting white-noise signal was scaled to obtain current fluctuations with a standard deviation of 8~pA. Injection of this input into the distal apical dendrite at a distance of 1094~$\si{\um}$ away from the soma of the active cell held at $-60$~mV yielded local membrane potential fluctuations with a standard deviation of 1.5~mV . The exact same white-noise input was used in all simulations (i.e., so-called "frozen" noise).

\subsection*{Quasi-active approximation of voltage-dependent ion currents}
Voltage-dependent membrane currents often behave in a near-linear fashion for small perturbations around a holding potential. This can be exploited by making linear approximations, so-called "quasi-active" models, of the nonlinear ionic currents \citep{Mauro1970, Koch1984, Hutcheon2000, Remme2014}. In this way one can reduce the parameter space while retaining key dynamical features of the system. Results in Figures 2--7 used quasi-active currents to simplify the original, nonlinear cortical pyramidal cell model (see above) and to allow for a systematic study of the effects of active conductances on the LFP.  

We here briefly describe the derivation of a quasi-active description of a single cellular compartment. The compartment includes one active current $I_w$ that depends on the membrane potential $V(t)$ and is described by $I_w(t)=\bar{g}_w w(t)(V(t) - E_w)$, with peak conductance $\bar g_w$, reversal potential $E_w$, and gating variable $w(t)$. The passive leak current is given by $I_L(t)=g_L (V(t) - E_L)$, with conductance $g_L$ and reversal potential $E_L$. Finally, the axial current, i.e., the net current entering or leaving to neighboring cellular compartments, is denoted by $I_{axial}(t)$. The voltage of the compartment then evolves according to
\begin{equation}
c_m \frac{dV(t)}{dt} = -\bar{g}_w w(t) \big( V(t) - E_w \big) - g_L \big( V(t) - E_L \big) + I_{axial}(t),
\end{equation}
where the term $c_m\, \tfrac{dV(t)}{dt}$ is the capacitive current with membrane capacitance $c_m$. The dynamics of the gating variable $w(t)$ is given by
\begin{equation}
\tau_w \big(V\big) \frac{{\rm d} w(t)}{{\rm d} t} = w_\infty \big(V\big) - w(t),
\end{equation}
with voltage-dependent activation time constant $\tau_w \big(V\big)$ and activation function $w_\infty \big(V\big)$.

The quasi-active description is obtained by linearizing $V$ and $w$ around resting-state values $V_R$ and $w_\infty(V_R)$, respectively, by means of Taylor expansions. Defining the variable $m(t) \equiv (w(t) - w_\infty(V_R)) / \frac{\partial}{\partial V} w_\infty(V_R)$, we can write the linearized equation describing the voltage dynamics of a single compartment:
\begin{equation}
c_m \frac{dV(t)}{dt} = - g_L \Big( \gamma_R \big(V(t) - V_R \big) + \mu m(t) \Big) + I_{axial}(t),
\end{equation}
where $\gamma_R \equiv 1 + \bar{g}_w w_\infty(V_R)/g_L$, i.e., the ratio between the total membrane conductance (at $V_R$) and the leak conductance. The parameter $\mu \equiv (\bar{g}_w/g_L) (V_R - E_w) \frac{\partial}{\partial V}w_\infty(V_R)$ determines whether the quasi-active current functions as a positive feedback (when $\mu<0$; i.e., a regenerative current) amplifying voltage deviations from the holding potential $V_R$, or as a negative feedback (when $\mu>0$; i.e., a restorative current) counteracting changes in the voltage. When $\mu=0$, the quasi-active current is frozen and functions as a static passive current (throughout the text we will refer to this as the "passive-frozen" case). The dynamics of the linear gating variable $m$ is described by,
\begin{equation}
\tau_w (V_R) \frac{{\rm d} m(t)}{{\rm d} t} = V(t) - V_R - m(t). 
\end{equation}
The description of the ionic currents in a single compartment is easily extended to a multi-compartmental model where each compartment can have its own set of parameters to describe the passive and quasi-active currents. 

For the simulations with a single linearized h-type current or persistent sodium current (see Figure~\ref{fig:linearization}), we kept the passive parameters, as well as the peak conductance and activation time constant (at the specified holding potential) of the relevant active current, the same as in the original detailed model.

To systematically study the effect of the cellular distribution of a quasi-active current on the LFP (see Figure~\ref{fig:systematic}), we used three different channel density distributions: (1) linearly increasing with distance from the soma, (2) linearly decreasing with distance from the soma, and (3) a uniform distribution. The slopes of the increasing (decreasing) distributions were set such that the most distal tip of the apical dendrite had a sixty-fold larger (smaller) density compared to that of the soma (in line with experimental estimates for $I_{\rm h}$ distributions: \citealt{Mishra2015, Lorincz2002, Nusser2009, Kole2006}), and the total membrane conductance of the quasi-active current (i.e., summed over all compartments) was the same as the total passive leak conductance $g_L$. The passive leak conductance was set uniformly to 50~$\si{\micro \siemens}$/cm$^2$ for all cases. For $w_\infty(V_R) = 0.5$ the distance-dependent quasi-active peak conductance was $\bar g_w(x) = 5.29 + 0.242 \, x$~$\si{\micro \siemens}$/cm$^2$ for the linear increase, and $\bar g_w(x) = 143 - 0.109\, x$~$\si{\micro \siemens}$/cm$^2$ for the linear decrease, where the distance $x$ was measured in~$\si{\um}$ and had a maximum of 1291~$\si{\um}$. Note that the three distributions had the same total quasi-active membrane conductance summed over the neuronal membrane.

The parameters $\mu$ and $\gamma_R$ vary along the cell for the non-uniform channel distributions. We introduced $\mu^* \equiv \mu(x) g_L /\bar g_w(x)$, such that $\mu^*= (V_R - E_w) \frac{\partial}{\partial V}w_\infty({V_R})$ is independent of the distribution of the quasi-active conductance and can be specified as a single constant. We used $\mu^* = -0.5$ for the regenerative conductance, $\mu^* = 0$ for the passive-frozen conductance, and $\mu^* = 2$ for the restorative conductance. For the uniform distribution this gives the same values as those used in \citet{Remme2011}, namely $\mu = -1,\ 0,\ 4$ for the regenerative, passive-frozen and restorative cases, respectively. The activation time constant $\tau_w(V_R)$ of the quasi-active conductance was set to 50~ms (unless specified otherwise) in order to have dynamics similar to the h-type conductance. The intracellular resistivity was $R_a=100\ \Omega\,$cm, and the specific membrane capacitance $c_m=$1~$\si{\micro \farad}$/cm$^2$.

\subsection*{Calculation of extracellular potentials}

Extracellular potentials recorded inside the brain are generated by transmembrane currents from cells in the vicinity of the electrode contact~\citep{Nunez2006}. The biophysical origin of the recorded signals is well understood in the context of volume conduction theory. Extracellular potentials originating from a simulated multi-compartmental neuron model can be computed by first obtaining the transmembrane currents $I_n(t)$ from each compartment $n$ at position $\vec r_n$. Next, the extracellular potential $\phi(\vec r, t)$ at position $\vec r$ resulting from these transmembrane currents can be calculated \citep{Holt1999, Linden2014}:
\begin{equation}
\phi(\vec r, t) = \frac 1{4 \pi \sigma} \sum_{n=1}^{N} I_n(t) \int
\frac{d \vec r_n}{|\vec r - \vec r_n|},
\end{equation}
where $\sigma$ is the conductivity of the extracellular medium. This corresponds to the so-called line-source formula assuming the transmembrane currents to be evenly distributed along the axes of cylindrical neural compartments, see \cite{Linden2014} for a detailed description.

All simulations and computations of the extracellular potentials were carried out using \texttt{LFPy} \citep{Linden2014}, an open-source Python package that provides an interface to \texttt{NEURON} \citep{Carnevale2006}. The time step of the neural simulation was 0.0625~ms. For all simulations the first 1000~ms was discarded to avoid initialization effects. All simulation code used to produce the figures in this study are available upon request.

\section*{Results}
To examine the contribution of subthreshold, active conductances on the extracellular potential we started by considering a previously published, detailed model of a layer 5 cortical pyramidal neuron endowed with a large variety of active conductances (\citealp{Hay2011}; see Methods).

Note that the term LFP commonly refers to the low-pass filtered version (below 300-500~Hz) of the extracellular potential (to filter away spikes). Since there is no spiking activity in the present use of the model, we simply refer here to the unfiltered version of the extracellular potential as the LFP.

\subsection*{Active subthreshold conductances shape the LFP}

\begin{figure*}[htbp]
\centering
\includegraphics[scale=0.93]{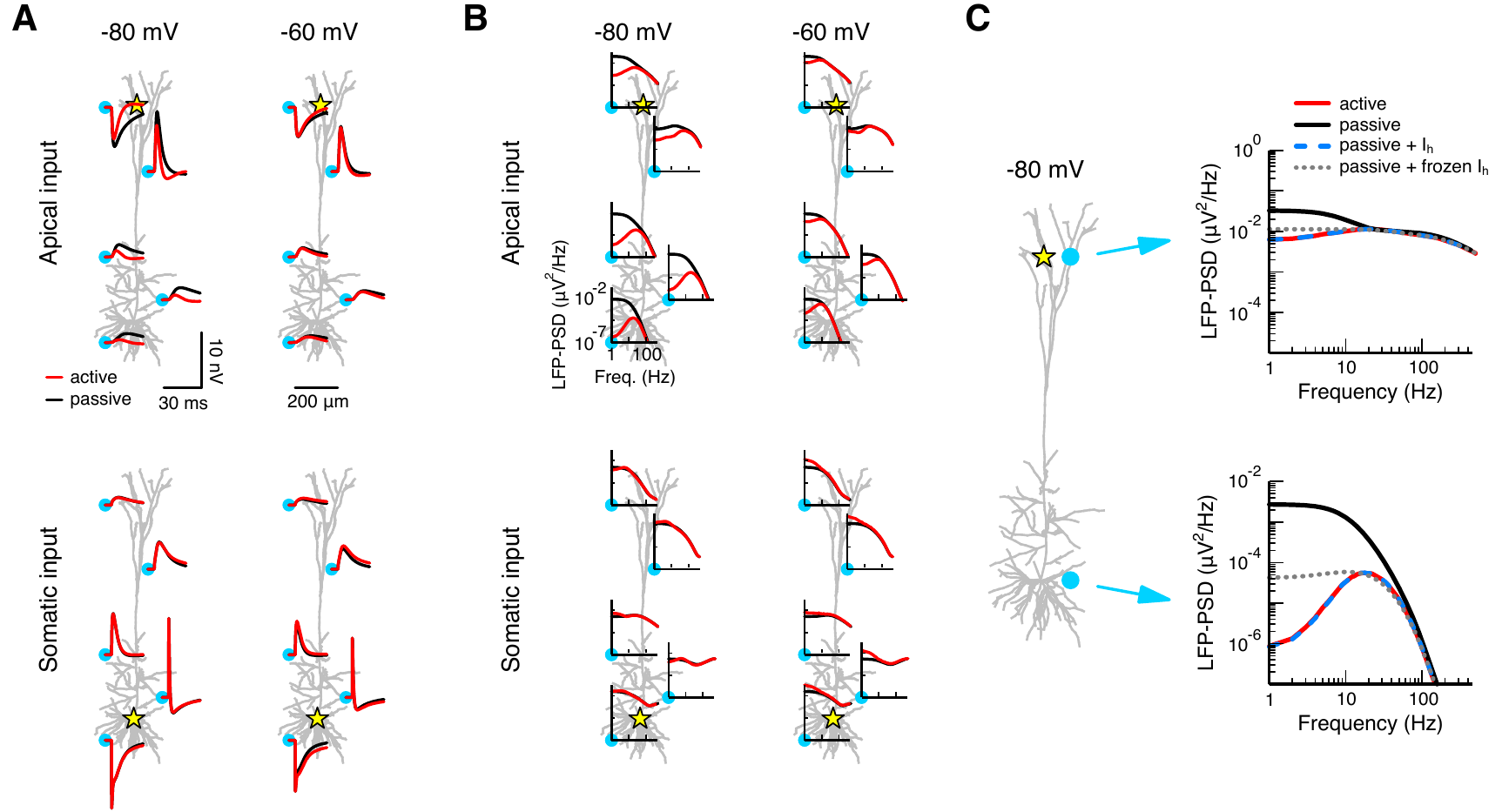}
\caption{\label{fig:intro} {\bf Active conductances can shape the extracellular signature of synaptic inputs.} 
\textbf{A}: A single synaptic input is provided to a cortical layer 5 pyramidal cell model. The extracellular response is shown at five positions (cyan dots) for two cases: the active model that includes various voltage-dependent conductances (red; see Methods), or a passive model from which the active conductances have been removed (black). The position of the input is marked by the yellow star: at the distal apical dendrite (top panels) or at the soma (bottom panels). The cell's resting potential was held uniformly at a hyperpolarized potential of -80~mV (left panels) or at a depolarized potential of -60~mV (right panels). The synaptic peak conductance was 0.001~$\si{\micro \siemens}$. The plots show the $x,z$-plane of the cell; the soma and the electrodes are positioned at $y$=0.
\textbf{B}:  As in panel A, but using white-noise current input (see Method section) instead of synaptic input, and displaying the response as the power spectral density (PSD). The PSD is calculated from 1000~ms long simulations.
\textbf{C}: Apical input (marked by star) to the cell model held at a hyperpolarized potential (-80~mV). The extracellular potential is shown at two positions (cyan dots) for the active (red) and passive (black) model and for two additional versions of the model: the passive model supplemented by $I_{\rm h}$ (dashed blue), and the passive model supplemented by frozen $I_{\rm h}$ (dotted gray; see Methods).}
\end{figure*}

We first determined the LFP in response to a single excitatory synaptic input. We compared the LFP signal of the active neuron model with a passive version of the model (i.e., all active conductances removed). The synaptic input was provided either to the distal apical dendrite (Figure~\ref{fig:intro}A, top row) or to the soma (Figure~\ref{fig:intro}A, bottom row) while the entire cell was hyperpolarized to $-80$~mV (Figure~\ref{fig:intro}A, left column) or depolarized to $-60$~mV (Figure~\ref{fig:intro}A, right column). The LFP was calculated at five different positions outside the neuron (marked by cyan dots). The difference between the LFPs of the active and passive models depended on the holding potential of the cell, the synaptic input position, as well as the extracellular recording position. In particular, we observed for hyperpolarized potentials that the active and passive LFPs were markedly different for apical synaptic input, while being very similar for somatic input.

In order to more fully characterize the effects of active currents on the LFP we next applied white-noise current input instead of a single synaptic input, and determined the power spectral density (PSD) of the resulting LFP (i.e., the square of the Fourier amplitude of the signal as a function of frequency). The white-noise input current had equal signal power at all integer frequencies in the considered frequency range (1--500~Hz; see also \citealp{Linden2010}). At most extracellular recording positions, a low-pass filtering was observed in the LFP of both the active and the passive model. This results from the intrinsic dendritic filtering (see, e.g., \citealp{Pettersen2012, Linden2010, Pettersen2008}) and increases with distance from the synaptic input site (Figure~\ref{fig:intro}B). The most striking difference between the active and the passive case was again found for the cell in a hyperpolarized state with apical input (Figure~\ref{fig:intro}B, top left), which showed a strong resonance (i.e., band-pass filter with a peak around 17--22~Hz) for the active model LFP, but not for the passive one. Note that the cell in a depolarized state and apical input shows qualitatively the same effect, though less prominent (Figure~\ref{fig:intro}B, top right).  

While the main effect of the active conductances in the case of apical input was to reduce the LFP power at the lowest frequencies, the effect from active conductances on somatic input was qualitatively different. For the cell with a depolarized resting potential, white-noise current resulted in an amplification of the low frequencies (Figure~\ref{fig:intro}B, bottom right). In all situations considered, a difference between the active and passive model was only apparent below about 30--50~Hz. At higher frequencies, capacitive currents dominated the transmembrane return currents both in the active and passive models~\citep{Linden2010}. The exact cross-over frequency where the capacitive currents become dominant over the ionic membrane currents, depends on neuronal properties (e.g., ion channel densities), as well as the positions of the input and the recording electrode.

Which ionic currents in the model are responsible for the large difference between the active and passive model for hyperpolarized potentials and with apical input? One obvious candidate is the hyperpolarization-activated inward current $I_{\rm h}$. This current operates at subthreshold voltages, is concentrated in the apical dendrites, and dampens the lowest frequency components of the membrane potential response to input \citep{Magee1998, Hu2009, Hu2002, Kole2006, Almog2014}. Indeed, we found that when we extended the passive model with only the h-type conductance, the LFP was indistinguishable from the full active model when the cell was in a hyperpolarized state and received apical input (Figure~\ref{fig:intro}C).
  
Two effects are involved in the reduction of the LFP signal power at low frequencies in the model with $I_{\rm h}$. First, with an additional active current comes an increased membrane conductance, which shifts the transmembrane return currents associated with the input current closer to the input position on the neuron. This gives smaller current-dipole moments and thus generally smaller LFPs \citep{Pettersen2008, Linden2010, Pettersen2012}. This is still a passive-membrane effect and can simply be incorporated into a passive model by rescaling the leak conductance corresponding to adding a "frozen" h-type conductance. The second effect stems from the dynamical properties of an active conductance such as $I_{\rm h}$, which actively counteract voltage fluctuations at low frequencies. While the extension of the passive model with a “frozen $I_{\rm h}$” was found to strongly dampen the low-frequency components of the LFP compared to the full active case (Figure~\ref{fig:intro}C, gray dotted lines), it failed to reproduce the resonance of the models with the dynamic $I_{\rm h}$ conductance (blue dashed and red). This demonstrates that the resonance observed in the LFP was an effect caused by the dynamics of the h-type channels, similar to previously reported $I_{\rm h}$-induced resonances in the membrane potential~\citep{Hutcheon2000, Hu2009, Zhuchkova2013}.

\subsection*{Linear models capture the effects of active conductances on the LFP in the subthreshold regime}

Above, we established that the h-type conductance is key in shaping the LFP for subthreshold input to the apical dendrites in the model by \citet{Hay2011}. Because in reality there is considerable variation between neurons regarding their biophysical properties, and perhaps not all relevant subthreshold currents were captured by this specific model, we next set out to perform a systematic study of the effects that any subthreshold active conductance can have on the LFP. For this we made use of the so-called "quasi-active" approximation of voltage-dependent currents. Active membrane conductances in general exhibit nonlinear behavior, but for small deviations of the membrane potential around a holding potential, active conductances often behave in a near-linear fashion. The dynamics of an active conductance can then be approximated by linearizing the current around a holding potential (\citealp{Mauro1970,Koch1984,Remme2011, Remme2014}; see Methods). This retains the voltage-dependent current dynamics for potentials not too far away from that potential, while strongly reducing the parameter space of the model and allowing an intuitive understanding of the model behavior.

Linearized active conductances can be divided into two classes: restorative and regenerative. Restorative conductances dampen the lowest frequency components of the membrane potential response to synaptic input, also leading to narrower synaptic voltage responses~\citep{Remme2011}. The intrinsic low-pass filtering of the cellular membrane~\citep{Koch1999}, in combination with the active dampening of low frequencies (i.e., high-pass filtering) by restorative conductances, can thus cause a resonance in the membrane potential PSD \citep{Remme2011, Zhuchkova2013, Hu2009}. Examples of restorative currents are the h-current and the M-type slow potassium current. In contrast, regenerative conductances amplify the low-frequency components of the membrane potential response to synaptic input, making the voltage responses to synaptic input broader in time as they propagate along the dendrites. This will only add to the intrinsic low-pass filtering of the cellular membrane, and as such, regenerative conductances lack the easily recognizable feature of the restorative conductances. Examples of regenerative conductances are the persistent sodium current $I_{\rm NaP}$ and the low-voltage activated calcium current~\citep{Remme2011}.

\begin{figure*}[htbp]
\centering
\includegraphics{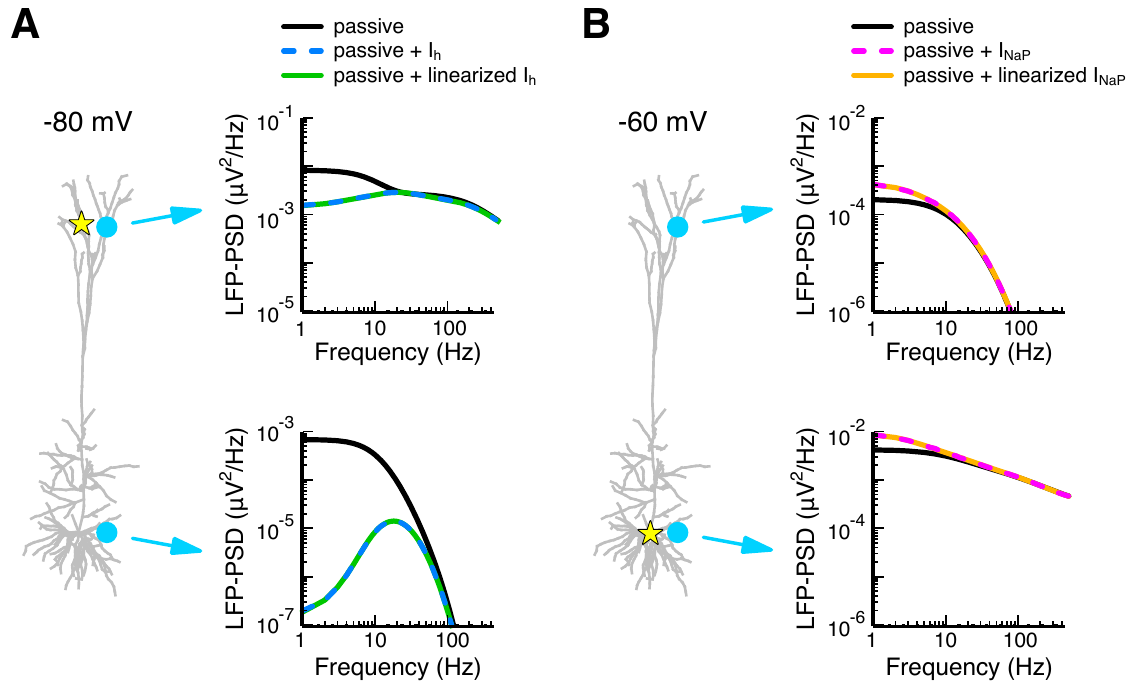}
\caption{\label{fig:linearization} {\bf Quasi-active conductances accurately mimic the responses of nonlinear conductances.}
\textbf{A:} Apical white-noise current input (yellow star) is provided to a cortical pyramidal cell model held at a hyperpolarized potential (-80~mV). The LFP-PSD is shown at two locations (cyan dots) for a passive model (black), a passive model with $I_{\rm h}$ (dashed blue), and a passive model that includes a linearized or quasi-active $I_{\rm h}$ (green; see Methods).
\textbf{B:} As in panel A, but with somatic white noise-current input and using the following three models: a passive model (black), a passive model with $I_{\rm NaP}$ (dashed magenta), and a passive model with linearized $I_{\rm NaP}$ (orange).
}
\end{figure*}

It is well established that quasi-active approximations of active currents can accurately capture the subthreshold effects on the membrane potential \citep{Koch1984, Remme2011}. To test whether this conclusion also holds for the LFP, we started with the original cortical pyramidal cell model from \citet{Hay2011}, and simplified it to two separate cases with only a single active conductance remaining in each. For the first case, we kept only the restorative $I_{\rm h}$ conductance, with its increasing conductance along the apical dendrite away from the soma. The resting potential was set uniformly to the hyperpolarized potential of $-80$~mV and the cell was stimulated with apical white-noise input. In the second case we kept only the regenerative persistent sodium current, solely present in the soma, set the resting potential uniformly to the depolarized potential of $-60$~mV, and applied somatic white-noise input. The $I_{\rm h}$-only model yielded the resonance in the LFP that was discussed above (Figure~\ref{fig:linearization}A, blue dashed lines), while the $I_{\rm NaP}$-only model gave an amplification of the low-frequency components of the LFP (Figure~\ref{fig:linearization}B, pink dashed lines). We then simulated the models with linearized versions of the original conductances and found the LFP from the active and linearized conductances to be indistinguishable (Figure~\ref{fig:linearization}, blue versus green, pink versus orange). In fact, for all tested combinations of resting potentials, input positions, and LFP recording positions, the quasi-active approximation was found to faithfully reproduce the signature of the active conductance in the membrane potentials, the transmembrane currents, as well as in the LFP. This applied as long as the perturbations of the membrane potential around the resting potential were not too large. For $I_{\rm h}$, fluctuations of 10--15~mV are typically within the appropriate range (data not shown, but see, e.g., \citealp{Remme2011}). We conclude that systematic studies with a single quasi-active conductance are justified.

\subsection*{Spatially asymmetric distributions of ion channels and inputs enhance the effect of active currents on the LFP}

\begin{figure*}
\centering
\includegraphics{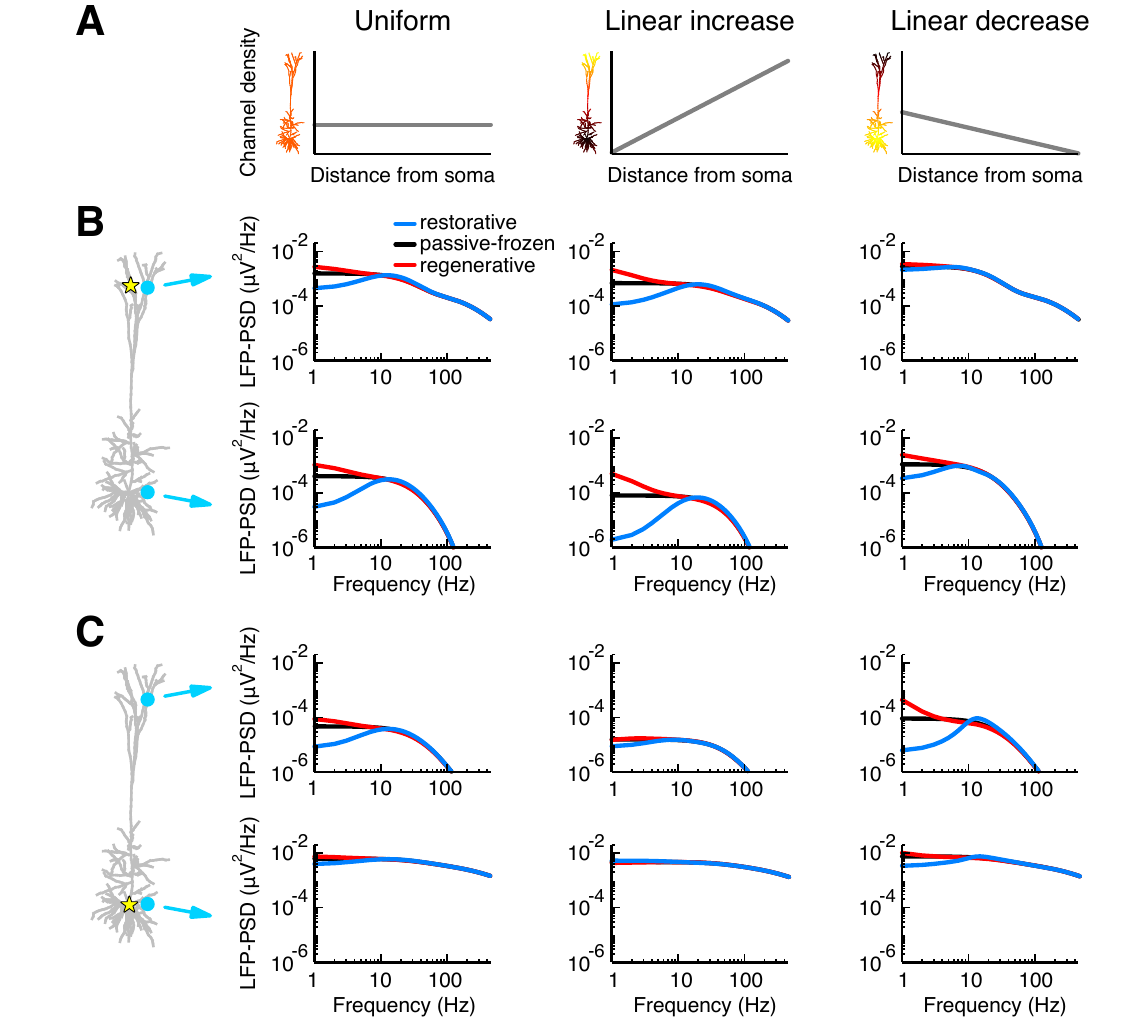}
\caption{\label{fig:systematic} {\bf The type of active conductance, its distribution across the cell membrane, and the location of the input shape the LFP}. \textbf{A:} A single quasi-active conductance was distributed across the pyramidal cell model in three ways: uniformly (left), linearly increasing (center), or linearly decreasing (right) with distance from the soma. The density of the linearly increasing conductance increased sixty-fold from the soma to the most distal apical dendrite, while the linearly decreasing conductance had a sixty-fold decrease over the same range.
\textbf{B:} Apical white-noise current input (yellow star) was provided to the cell model and the LFP-PSD was determined at two positions (cyan dots). The quasi-conductance was either regenerative (red), passive-frozen (black) or restorative (blue). The three cases had the same total resting membrane conductance and the total passive leak conductance was equal to the total quasi-active conductance.
\textbf{C:} As panel B, but with white-noise current input provided to the soma.
}
\end{figure*}

We next turned to a systematic exploration of how active subthreshold currents affect the LFP. For this we used a pyramidal cell model with a single quasi-active current. The key parameter characterizing the quasi-active current is $\mu^*$, the sign of which determines whether the current is regenerative ($\mu^*<0$) or restorative ($\mu^* > 0$), while  $\mu^*=0$ for a passive-frozen conductance (i.e., the added conductance has no dynamics and thus simply adds to the passive leak conductance). The quasi-active current was distributed across the cell either uniformly, linearly increasing with distance from soma, or linearly decreasing with distance from soma (Figure~\ref{fig:systematic} -- left, middle and right columns, respectively). In all cases the models had the same total membrane conductance summed over the neuronal membrane. We applied white-noise current input either to the apical dendrite (panel B) or to the soma (panel C) and calculated the LFP-PSD for each case in the apical region and in the somatic region. A large difference between the results from the passive-frozen model and the restorative or regenerative models would imply that the dynamics of the active conductance in question shapes the LFP. Indeed, we found for most scenarios, that the regenerative current caused an amplification of the low-frequency components of the LFP (Figure~\ref{fig:systematic}, red lines) compared to the passive-frozen case (black line), while a restorative current dampened the low frequencies, leading to the previously discussed resonance (Figure~\ref{fig:systematic}, blue lines). However, as discussed below, there were exceptions. 

We found that the impact of the quasi-active current was always largest in the LFP recorded on the opposite side of the cell compared to the position of the input, i.e., next to the soma for apical input (panel B, bottom row) and next to the apical dendrite for somatic input (panel C, top row). This can be understood since the LFP is a distance-weighted sum of all transmembrane currents, so that transmembrane currents close to the recording electrode will contribute more to the measured LFP signal. For recordings on the opposite side relative to the input, the transmembrane return currents have been strongly affected by the active conductance after propagating along the apical dendrite. The effect was also largest when the quasi-active current was non-uniformly distributed and concentrated at the position of the input (panel B, middle column; panel C, right column), and smallest when the input was to a region of low density (panel B, right column; panel C, middle column). This also follows from the reasoning above: when the input is to a region with a low density of an active current, the return currents close to the input will be hardly affected by the active current. As the signal propagates along the dendrites to regions with higher densities of the active current, it will be increasingly affected by the active current, however, the signal amplitude will also gradually decrease. Therefore, the LFP is less affected by the presence of a non-uniformly distributed active current when the input is provided to a region with low densities compared to when input is provided to a high-density region.

\begin{figure*}[t]
\centering
\includegraphics{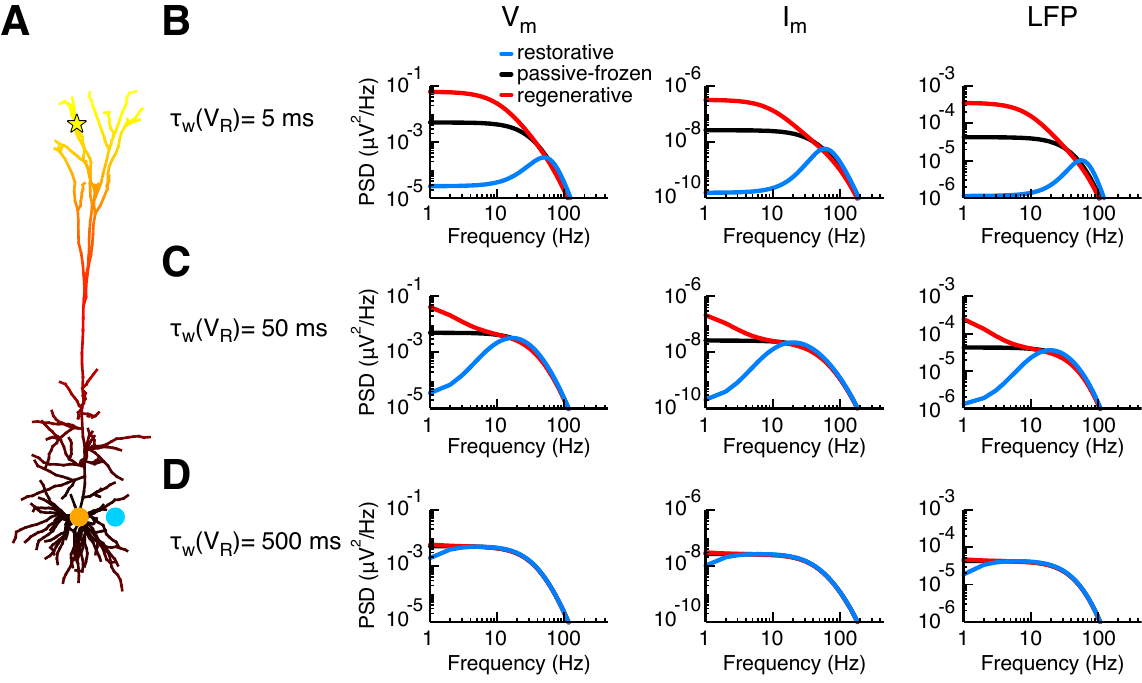}
\caption{\label{fig:timeconstant} {\bf Activation time constants of active conductances determine the frequencies for which they affect the LFP.}
\textbf{A}: Apical white-noise current input (yellow star) to the cortical pyramidal cell model with a single linearly increasing quasi-active conductance (as in middle column of Fig. \ref{fig:systematic}B). 
\textbf{B}: The quasi-active conductance is either restorative (blue), passive-frozen (black) or regenerative (red), and the PSD is shown for the somatic membrane potential $V_m$ (left panel), the somatic transmembrane currents $I_m$ (middle panel), and the LFP (right panel) at a position close to the soma (cyan dot). The activation time constant of the quasi-active conductance is $\tau_w(V_R) = 5$~ms.
\textbf{C}: As panel B, but with $\tau_w(V_R) = 50$~ms.
\textbf{D}: As panel B, but with $\tau_w(V_R) = 500$~ms.
}
\end{figure*}

Besides the type of an active current (i.e., restorative or regenerative) and how it is distributed across the neuron, also its activation time constant is key for the effects on the LFP. The dynamics of an active current will only express its effect on the membrane potential, transmembrane currents and LFP for frequencies that are sufficiently low relative to the activation time constant ($\tau_w (V_R)$). This is seen by considering different activation time constants for the quasi-active conductance (Figure~\ref{fig:timeconstant}). Increasing the activation time constant decreased the frequency at which the regenerative/restorative responses became similar to the passive-frozen case. For the resonance this means that its peak becomes broader and shifts to lower frequencies for larger activation time constants.

\subsection*{Resonance is expressed for asymmetrically distributed synaptic input}

In the above simulations we considered a single white-noise current input to characterize the effect of active currents on the LFP. We next examined whether the observed effects also hold when the cell receives a barrage of synaptic input as is typical for \emph{in vivo} conditions \citep{Destexhe2003}.

\begin{figure*}[t]
\centering
\includegraphics{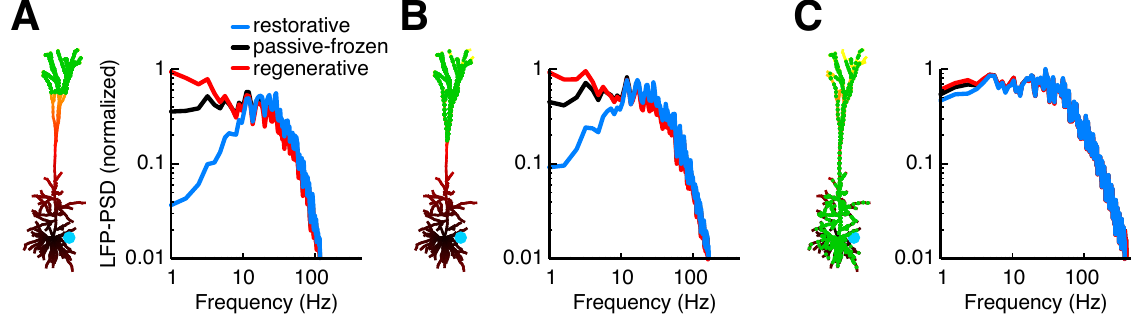}
\caption{\label{fig:distributed_synaptic} 
{\bf Resonance is retained for asymmetric input from distributed synapses.} 
\textbf{A}: The pyramidal cell model expressed a single linearly increasing quasi-active conductance (see Figure~\ref{fig:timeconstant}A) that was either restorative (blue), passive-frozen (black), or regenerative (red). 1000 excitatory conductance-based synapses (green dots) with a peak conductance of 0.0001~$\si{\micro \siemens}$ were distributed across the distal apical tuft more than 900~$\si{\um}$ away from soma. The synapses were activated by independent Poisson processes with a mean rate of 5 spikes per second. Simulations were run for 20 seconds and the LFP-PSD was calculated using Welch's method.
\textbf{B}: As panel A, but with synapses distributed above the main bifurcation, 600~$\si{\um}$ away from the soma.
\textbf{C}: Synapses were distributed uniformly across the entire cell.
}
\end{figure*}

We considered a cell model with a single quasi-active conductance that increases linearly in density with distance from the soma. The model received input from 1000 excitatory synapses that were distributed across the neuronal membrane and randomly activated according to independent Poisson statistics. We considered three synaptic distributions: (1) uniformly distributed across the distal apical tuft dendrites located more than 900~$\si{\um}$ from the soma (Figure~\ref{fig:distributed_synaptic}A), (2) uniformly distributed across the apical dendrites more than 600~$\si{\um}$ from the soma (Figure~\ref{fig:distributed_synaptic}B), and (3) uniformly across the entire cell (Figure~\ref{fig:distributed_synaptic}C). For each distribution the exact same 1000 synapse positions and input spike trains were used in the simulations for the regenerative, passive-frozen, and restorative conductances. Hence, the only difference between the simulations with a given synapse distribution was the nature of the quasi-active conductance. 

The resonance for the restorative model was clearly retained for LFP signals recorded close to the soma when the synaptic inputs were distributed asymmetrically, being concentrated in the apical dendrites (Figure~\ref{fig:distributed_synaptic}A,B). However, for uniformly distributed input (Figure~\ref{fig:distributed_synaptic}C) the resonance was practically undetectable. Similarly, for the regenerative model, increases of low frequencies in the LFP-PSD were only reliably observed for the models with asymmetric synaptic input distributions (Figure~\ref{fig:distributed_synaptic}A,B). Note that we obtained the same results when keeping the synapse densities fixed across the three synapse distributions, instead of keeping the total synapse number constant (results not shown). The dependence of the modulation of the LFP on the input distribution can be understood by considering the results in Figure~\ref{fig:systematic}: the case with uniform input approximately corresponds to a linear combination of the LFPs stemming from apical and somatic input (Figure~\ref{fig:systematic}, middle column). In the somatic region the net LFP signal will be strongly dominated by the transmembrane currents resulting from the somatic input, which are hardly affected by the active conductances since these are concentrated in the distal apical dendrites.

\subsection*{Resonance is a spatially stable feature in the LFP}

The LFP signals measured in \emph{in vivo} experiments are generated by many neurons located within a distance from a few hundred micrometers up to a few millimeters away from the recording electrode~\citep{Linden2011,Leski2013}. An effect of a quasi-active conductance on the LFP generated by a single neuron will only carry over to the LFP generated by an entire population if the effect is spatially "robust". For example, the resonance of the single-neuron LFP will only be observed in the population-LFP if it is present in a sizable part of the volume surrounding the cell. If it would only be recorded at special electrode positions, then it would be expected to average out when considering an entire population. The linearity of the LFP signal generation implies that the summed contributions of many single neuron-LFP signals with similar shape, would result in a population LFP with a similar shape. It is therefore key to study how the LFP varies in the space around the neuron.

\begin{figure*}
\centering
\includegraphics{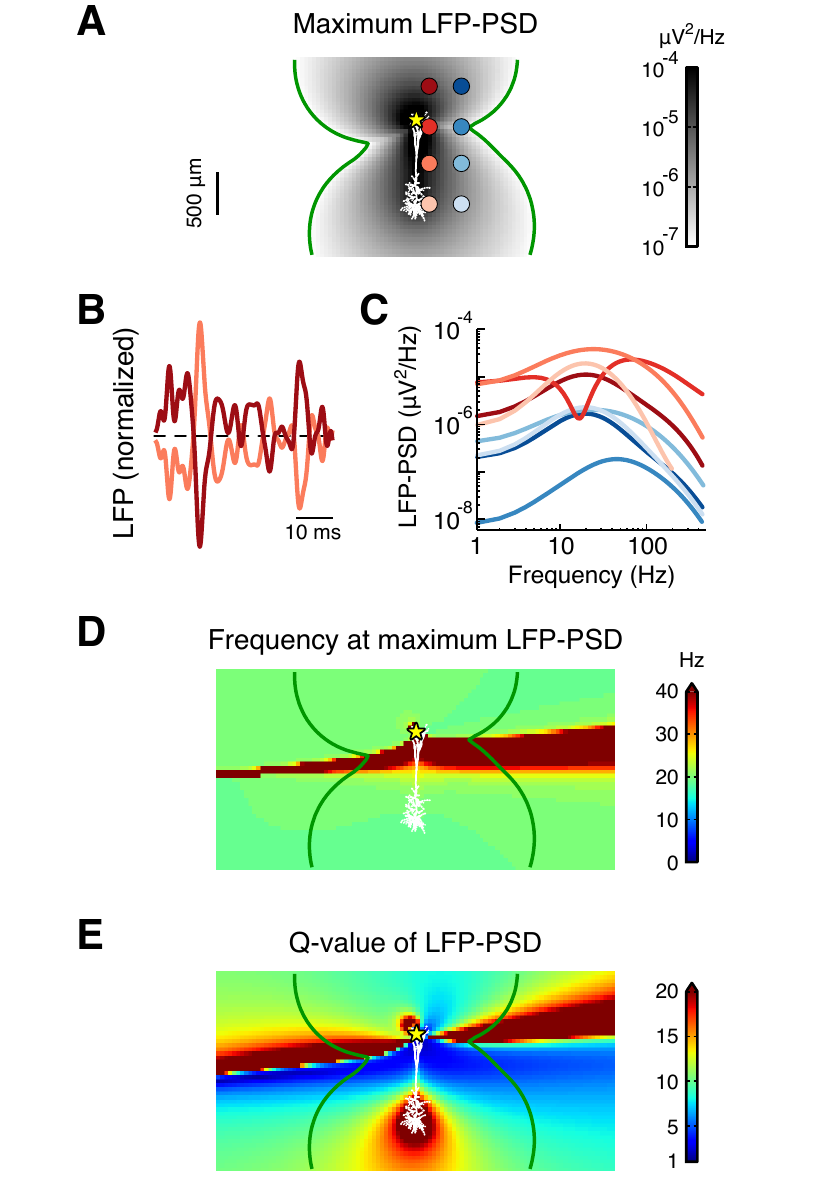}
\caption{\label{fig:LFP_with_distance} {\bf Resonance in the LFP resulting from a restorative conductance is a spatially stable feature.}
\textbf{A}: Apical white-noise current input (yellow star) is provided to the cortical pyramidal cell model with a single quasi-active conductance with increasing density with distance from soma. The resulting LFP-PSD was calculated at a dense 2D-grid surrounding the cell. The maximum power of the LFP is denoted in shades of gray. The green contour line shows where the LFP power is $10^{-7}\ \si{\uV}^2$/Hz.
\textbf{B}: Excerpts of LFP time traces at two extracellular positions, corresponding to the two electrodes at the opposite side of the zero crossing region of the dipole in panel A (i.e., the first and third electrode contact in the left column).
\textbf{C}: LFP-PSD computed at the eight electrode positions marked by shades of red and blue in panel A.
\textbf{D}: The frequency for which the LFP-PSD shows the maximum power (see panel A).
\textbf{E}: The Q-value of the LFP-PSD, defined as the maximum power (see panel A) divided by the power at 1~Hz.
}
\end{figure*}

We calculated the LFP at a dense grid of positions around a pyramidal cell receiving apical white-noise input (Figure~\ref{fig:LFP_with_distance}). The model expressed a restorative quasi-active conductance that increased in density with distance from the soma. Our previous results (see Figure~\ref{fig:systematic}) demonstrated that such a model can show strong resonances in the LFP. The maximum amplitude of the LFP-PSD on the position-grid around the cell had the shape of a dipole (panel A), as expected following the requirement of charge conservation~\citep{Linden2010}. This dipolar nature of the LFP pattern was also demonstrated by comparing two normalized LFP time traces at opposite sides of the zero-crossing region of the dipole and noting that they are almost mirror images (panel B; depicted traces correspond to first and third electrode contacts from the top in the first column in panel A).

For two columns of electrode positions (marked in panel A) we examined the LFP-PSD in detail (panel C). The LFP signals again demonstrated that the resonance was most pronounced on the side of the cell opposite to the input (compare lightest trace to the darker ones). Note that the electrodes positioned in the zero-crossing region of the dipole exhibited more variable PSD shapes, for example, a strong band-stop at around 20~Hz. This is because different frequency components of the LFP will have slightly different zero-crossing regions (see also~\citealt{Linden2010}). We focused on the LFP resonance and determined the frequency of maximum LFP power for the grid of positions around the neuron. The peak frequency was found to be stable at around 20~Hz, except in the zero-crossing region of the dipole (panel D).

The so-called Q-value is commonly used to characterize resonances. In the neurophysiological literature it is often defined as the magnitude of a variable (typically the membrane potential) at the resonance frequency divided by its value at the lowest frequency considered \citep{Hutcheon1996}. We found that the LFP Q-value was stable and large at the soma region of the cell (panel E), i.e., on the side opposite to the input, in accordance with observations from Figures~\ref{fig:intro}--\ref{fig:systematic}.
Note that the largest LFP Q-values were in fact observed in the zero-crossing region. However, in this region it does not reflect a resonance caused by a restorative current, but rather the very variable LFP pattern in this region. The lower signal power in this region (Figure~\ref{fig:LFP_with_distance}A), in combination with the fact that the position of the zero-crossing region will depend on various parameters, such as the exact position of the synaptic input, means that the very large Q-values seen here will not carry over to the population LFP.

\subsection*{Spatial profile of the membrane potentials, transmembrane currents and LFP}

Above we demonstrated an amplification and a dampening of low-frequency LFP components caused by regenerative and restorative conductances, respectively. These effects were qualitatively similar to the effects of such active conductances on the membrane potential as reported by, for example, \citet{Remme2011}. However, as LFP signals reflect transmembrane currents, it is a fundamentally different measure of neural activity than the membrane potential~\citep{Pettersen2012,Einevoll2013}. It can therefore not be expected \emph{a priori} that quasi-active conductances always have qualitatively similar effects on the LFP and the membrane potential. We thus next studied the differences in the effects of active currents on the membrane potential, transmembrane currents and LFP in a simplified setting, i.e., white-noise current input injected into a single, long neurite with uniform passive membrane parameters and a single uniformly distributed quasi-active conductance.

\begin{figure*}[htbp]
\centering
\includegraphics{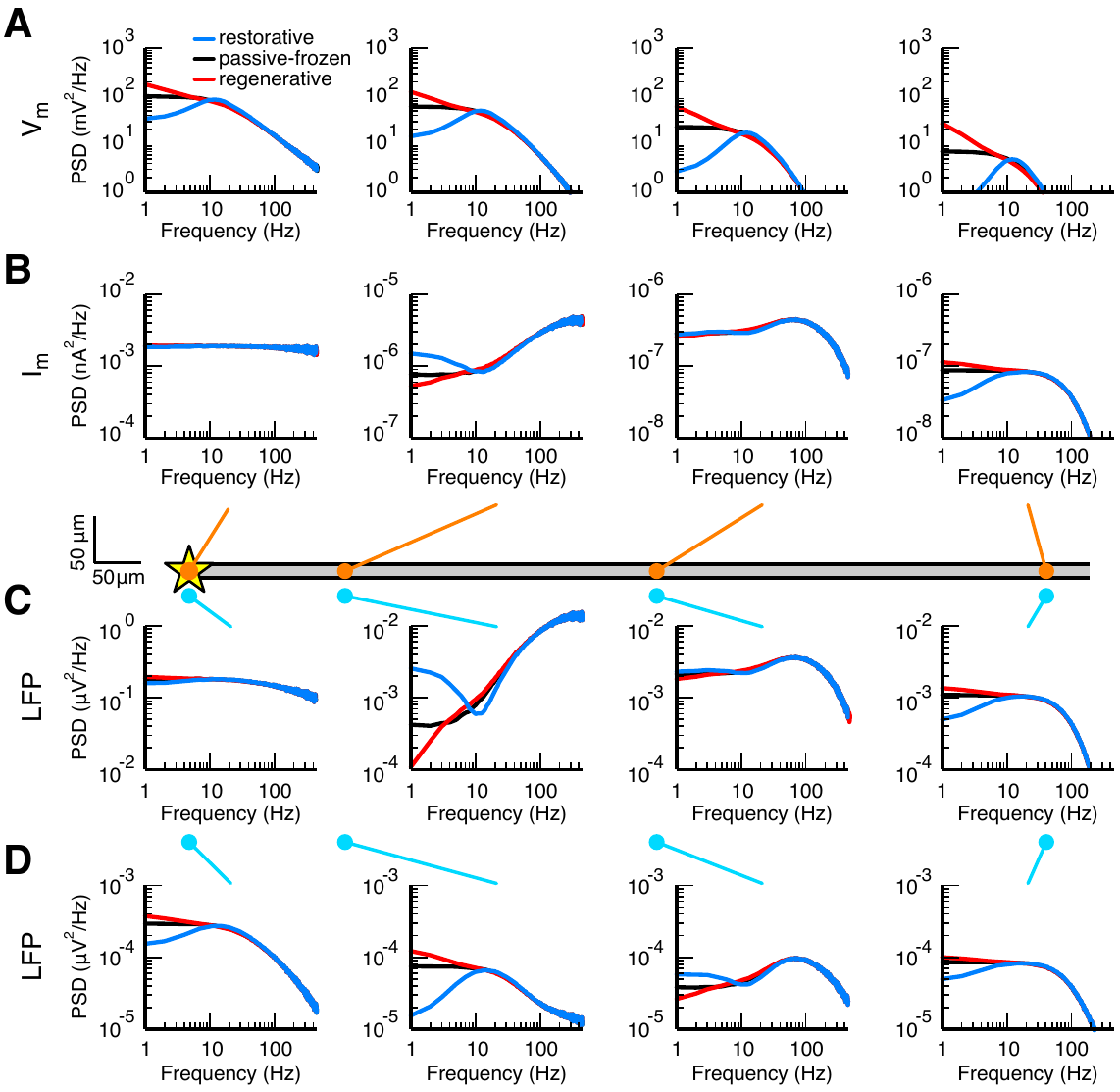}
\caption{\label{fig:inf_neurite} {\bf Voltage, transmembrane current, and LFP for a semi-infinite neurite with a quasi-active conductance.}
White-noise current input (yellow star in schematic above panel C) was applied to the end of a semi-infinite neurite with a single quasi-active conductance that was either restorative (blue), passive-frozen (black) or regenerative (red). The neurite had length 2000~$\si{\um}$, diameter 2~$\si{\um}$, intracellular resistivity $R_a=100\ \Omega\,$cm, and the passive leak and quasi-active conductance density were both uniformly set to 50~$\si{\micro \siemens}$/cm$^2$.
\textbf{A}: The membrane potential PSD ($V_m$) is shown at the positions marked by the orange dots in the schematic (at 0, 177, 531 and 973~$\si{\um}$ distance from the input). 
\textbf{B}: As in panel A, but showing the transmembrane current PSD at the positions of the orange dots.
\textbf{C}: As in panel A, but showing the LFP-PSD for the positions marked by the cyan dots (20~$\si{\um}$ away from the neurite). 
\textbf{D}: As in panel C, but showing the LFP-PSD for electrodes at 300~$\si{\um}$ away from the neurite, marked by the cyan dots.
}
\end{figure*}

Focusing first on the membrane potential ($V_m$, Figure~\ref{fig:inf_neurite}A), we indeed observe that the low-frequency components are always amplified by regenerative currents (red), whereas they are always dampened by restorative conductances (blue), compared to the passive-frozen case (black). The effects increase with "exposure" to the quasi-active current, i.e., how far the input has propagated along the neurite \citep{Remme2011}. In addition, the membrane potential is increasingly low-pass filtered with distance, resulting in a stronger membrane-potential resonance with distance from the input site.

The position dependence of transmembrane currents is more complicated ($I_m$, panel B). We first consider the passive-frozen case: at the input site (panel B, leftmost) the transmembrane current is approximately equal for all frequencies, as expected due to the white-noise current input. From charge conservation it follows that the current that enters the cell must also exit the cell, i.e., that transmembrane currents across the neuron sum to zero~\citep{Koch1999}. Since high-frequency components propagate less than low-frequency components, they dominate close to the input. Hence, the transmembrane currents show a high-pass filtered PSD at positions close to the white-noise input (panel B, second column). The reverse is true far away from the input (panel B, rightmost), where there is a low-pass effect similar as for the membrane potential. This results from the stronger propagation of the low frequencies, which therefore dominate over the high frequencies at larger distances~\citep{Koch1999,Linden2010}. Note that at intermediate distances, a “passive resonance” (panel B, third column) can be observed through the combination of a low-pass and a high-pass filter.

A restorative current functions as a negative feedback, counteracting membrane potential changes, such that voltage signals do not spread as far down the neurite as in the passive-frozen case (panel A; compare blue and black traces). Therefore the return currents resulting from the current injections move closer to the input site, leading to an amplification of the transmembrane currents close to the input site, and a decrease far away from the input site, relative to the passive-frozen case (panel B).

Regenerative currents amplify membrane potential deviations along the neurite (panel A; compare red and black traces) and thus shift the return currents further away from the input compared to the passive-frozen case. This leads to a decrease in the transmembrane currents close to the input site, and an amplification far away from the input site compared to the passive-frozen case (panel B).

Because the LFP signal is a distance-weighted sum of all transmembrane currents, the described effects in the transmembrane currents are all present in the LFP measured very close to the neurite (panel C; 20~$\si{\um}$). For electrodes further away from the neurite (panel D; 300~$\si{\um}$) the LFP is no longer dominated by transmembrane currents from any single part of the neurite. Since the restorative currents move the transmembrane return currents closer to the input site, a smaller current dipole moment is generated, thus resulting in a decrease of the LFP-power at most recording positions. The opposite is found for the model with a regenerative conductance. As we demonstrated in Figure~\ref{fig:timeconstant}, these effects hold for the low-frequency components of the LFP, because the dynamics of the active current only affect the cell response for frequencies that are slow compared to the activation time constant.

The resulting transmembrane currents from a single white-noise input into a neurite also helps explain why we needed asymmetric input to retain the effects that an active current has on the LFP-PSD (see Figure~\ref{fig:distributed_synaptic}). For a single input location, the effects of an active current on the low-frequency components of the transmembrane currents close to, and far away from the white-noise input were opposite (Figure~\ref{fig:inf_neurite}B, second versus fourth panel). Hence, for input that is localized to one region of a cell, the contributions from active currents will sum "constructively" and largely retain the characteristic LFP signatures for a single individual synaptic input. For uniform input, however, an active current will give a combination of amplifying and dampening effects which will sum "destructively", decreasing the total effect of the active current on the shape of the LFP.

\section*{Discussion}

In the present paper we explored the role played by active conductances in shaping the local field potential (LFP) generated by neurons. It thus goes beyond previous principled studies on the LFP generated by synaptically activated passive neurons~\citep{Linden2010,Linden2011,Leski2013}. While the various effects of active conductances in forming the LFP have previously been explored in comprehensive network simulations of thousands of spiking neurons~\citep{Reimann2013}, the present work was instead tailored to probe subthreshold effects in detail.

With a morphologically detailed and experimentally constrained cortical pyramidal cell model, we demonstrated that active conductances can strongly shape the LFP stemming from subthreshold input. The precise effects depended on the position of the input, the position of the extracellular electrode, and the membrane potential of the cell (Figure~\ref{fig:intro}). The subthreshold active conductances had distinct effects on the power spectral density (PSD) of the LFP, providing either amplification or attenuation of the low frequencies, in the latter case leading to a resonance in the LFP-PSD. The effects on the LFP were observed for various spatial distributions of the active conductances across the cell, but generally required an asymmetric distribution of synaptic input onto the cell (Figure~\ref{fig:distributed_synaptic}). We found that the effect of a subthreshold active conductance on the LFP was maximized for (i) an asymmetric distribution of the active conductance, (ii) when the input was targeted to regions where the active conductance was most strongly expressed, and (iii) the LFP was recorded on the opposite side of the cell with respect to the input (Figure~\ref{fig:systematic}). 

Peaks in the LFP-PSD are commonly observed experimentally (see, e.g.,~\citealp{Roberts2013, Hadjipapas2015}), and such peaks are usually interpreted as the result of network oscillations, i.e., oscillatory firing activity driving the LFP-generating neurons~\citep{Buzsaki2004}.
Importantly, our work shows that such peaks may also be due to subthreshold restorative conductances molding the transmembrane return currents. In particular, in our simulations with the pyramidal cell model by \citet{Hay2011} we found that the h-type current had a prominent role in shaping the LFP and could cause a strong resonance in the LFP-PSD (Figures~\ref{fig:intro}C, \ref{fig:linearization}A). The h-type current is strongly expressed in cortical and hippocampal pyramidal neurons and has a particularly asymmetric distribution across the cell, increasing in density along the apical dendrites and peaking in the distal apical tuft dendrites \citep{Magee1998, Williams_2000, Harnett2015}. The h-type current can be expected to impact the LFP resulting from synaptic input that is predominantly targeting the apical dendritic tuft of populations of cortical or hippocampal pyramidal neurons. Indeed, various input pathways to pyramidal neurons target specific domains of the cell (see, e.g., \citealp{Petreanu2009}). An intriguing consequence of our work is that the LFP can contain information on the spatial distribution of subthreshold active channels, as well as on the location of synaptic input to the LFP generating cells. For example, given a known apical concentration of h-type conductances, a resonance in the LFP due to $I_{\rm h}$ would suggest that the cells receive asymmetric input. Vice versa, if asymmetrical input is provided, an absence of a resonance in the LFP (measured at multiple locations along the cell axis) suggests that the cell does not strongly express restorative currents.

A direct approach to test our model predictions would be through {\it in vitro} cortical slice experiments. The asymmetric input can be provided by targeting the apical dendrites using, e.g., glutamate uncaging. The polarized distribution of $I_{\rm h}$ that pyramidal neurons typically display, should then give rise to a notable resonance in the LFP-PSD, in particular when measured close to the pyramidal cell soma (Figure~\ref{fig:LFP_with_distance}). Addition of an $I_{\rm h}$ blocker (e.g., ZD7288) is expected to remove any resonance caused by $I_{\rm h}$.

In most of our simulations we described voltage-dependent channels with so-called quasi-active currents, which are linear approximations of the voltage-dependent currents \citep{Mauro1970, Koch1984, Hutcheon2000, Remme2014}. The linear descriptions highlights that there are basically two types of active currents: regenerative and restorative, which have opposite effects on the voltage response, amplifying or counteracting voltage deflections, respectively. The quasi-active description not only permits the use of various analytical techniques for linear systems, but also strongly reduces the number of parameters to describe an active current. This simplified description enables the systematic study of the effect of subthreshold input on the voltage response of active, dendritic neurons and allows a thorough understanding of the dynamics of the system \citep{Koch1984, Bressloff1999, Coombes2007, Goldberg2007, Remme2009, Remme2010, Remme2011}. We showed here that the quasi-active current descriptions can also be used to accurately capture the LFP signal (Figure~\ref{fig:linearization}). Because the h-type current was dominating the subthreshold effects on the LFP in the model by~\citet{Hay2011}, a single quasi-active current was sufficient to approximate the LFP signal of the full, nonlinear model. The approach enabled us to establish an overview of the effects of active currents on the LFP signal. We achieved this by using the detailed neuron morphology with a single quasi-active current, where the current was either restorative or regenerative, had a range of activation time constants (Figure~\ref{fig:timeconstant}), and was distributed in various ways across the cell (Figure~\ref{fig:systematic}).

In the \emph{in vivo} situation, the measured LFP will reflect the activity in populations of neurons~\citep{Einevoll2013}. A natural extension of the present work will be to investigate the effect of active dendritic currents on the LFP in a population of neurons receiving synaptic inputs~\citep{Linden2011,Leski2013}. For the case of passive dendrites the key factor in determining the magnitude and spread of the population LFP was found to be the amount of correlation between the numerous synaptic inputs driving the population~\citep{Linden2011,Einevoll2013,Leski2013}. The synaptic correlation level will be expected to also be key for determining the population LFP in the case of active dendritic conductances, but insight into its specific influence will require detailed modeling studies. A key simplifying feature of the present study was the observation that linearized quasi-active conductances accounted very well for the salient effects of active subthreshold conductances on the single-neuron LFP. This approximation will also be applicable when studying effects of subthreshold active conductances on population LFPs, thus assuring linearity of the model system and greatly simplifying the analysis as in ~\citet{Linden2011} and \citet{Leski2013}.

In the present study we have not considered suprathreshold active conductances, i.e., the active channels underlying spike generation. While the present biophysical modeling scheme is equally applicable to spikes~\citep{Holt1999,Gold2006,Pettersen2008}, the contributions to the LFP from such spikes may be expected to be negligible for cortical networks in the \emph{in vivo} situation, at least for the low frequencies of LFP.
In \citet{Pettersen2008a} the extracellular potential generated by a synaptically-activated population of 1040 pyramidal neurons mimicking a layer-5 population in rat somatosensory (barrel) cortex was modeled by the present biophysical scheme. In order for the model predictions to be in accordance with measured extracellular potentials from the rat barrel cortex, both for the LFP (frequencies $\lesssim$ 500~Hz) and for multi-unit activity (frequencies $\gtrsim$ 500~Hz), only 4\% (40 of 1040) of the model neurons were tuned to fire an action potential following the stimulus input. In this situation the modeled LFP signal was observed to be completely dominated by the synaptic input currents and their associated return currents, with negligible contributions from the spikes themselves. However, in situations with strongly correlated firing, for example during sharp-wave ripples in hippocampus, the spikes have been found to have sizable contributions to the extracellular potentials for frequencies down to 100~Hz~(\citealp{Schomburg2012}; see also \citealp{Reimann2013, Taxidis2015}). For further discussion on the biophysical origin of the extracellular potentials in this "high-gamma" range, see for example \citet{Ray2011} and \citet{Scheffer2013}.
Note also that as the extracellular potential is given as a linear sum of contributions from the various transmembrane currents surrounding the contact, putative contributions from spikes can be estimated by adding computed spike signatures~\citep{Holt1999,Gold2006,Pettersen2008} to the present results for the LFP.

Even though the LFP has been measured for more than half a century, the interpretation of the recorded data has so far largely been qualitative~(for a review, see \citealp{Einevoll2013}). We believe that the  present investigation on the role of active dendritic conductances in shaping the recorded signal, will turn out to be an important contribution towards the goal of making combined modeling and measurement of the LFP signal a practical research tool for detailed probing of neural circuit activity.

\subsubsection*{Competing Interests}
The authors declare no competing financial interests.
\subsubsection*{Funding}
The research leading to these results has received funding from the European Union Seventh Framework Program (FP7/2007-2013) under grant agreement 604102 (Human Brain Project, HBP), the Research Council of Norway (NevroNor, Notur, nn4661k), the Norwegian node of the International Neuroinformatics Coordinating Facility (INCF, NFR 214842/H10), and the Einstein Foundation Berlin.

\subsubsection*{Author contributions}
All authors took part in designing the work, interpreting data, and writing the manuscript. TVN wrote the simulation code. TVN and MWHR made the figures. All authors approved the final version of the manuscript and agree to be accountable for all aspects of the work in ensuring that questions related to the accuracy or integrity of any part of the work are appropriately investigated and resolved. All persons designated as authors qualify for authorship, and all those who qualify for authorship are listed.

\newpage

\clearpage
\bibliographystyle{jphysiol}
\bibliography{aLFP_papers}

\newpage

\clearpage

\clearpage

\end{document}